\documentclass[prd,twocolumn,nobibnotes,amssymb,preprintnumbers]{revtex4}
\begin{document}
\title{On Boundary RG-Flows in Coset Conformal Field Theories}
\author{Stefan Fredenhagen}
\email{stefan@aei.mpg.de}
\author{Volker Schomerus}
\email{vschomer@aei.mpg.de}
\affiliation{Max-Planck-Institute for Gravitational Physics\\
Am M{\"u}hlenberg 1, D-14476 Golm, Germany}
\date{May 1, 2002}
\preprint{AEI-2002-035}\preprint{hep-th/0205011}
\begin{abstract}
We propose a new rule for boundary renormalization group flows 
in fixed-point free coset models. Our proposal generalizes 
the 'absorption of boundary spin'-principle formulated by 
Affleck and Ludwig to a large class of perturbations in boundary 
conformal field theories. We illustrate the rule in the case
of unitary minimal models. 
\end{abstract}
\maketitle
Renormalization group (RG) flows in models with boundaries or 
defects are of interest in condensed matter theory, statistical 
physics, and in string theory where they describe aspects of 
D-brane dynamics. There exist various tools to investigate 
flows generated by boundary fields including the Thermodynamic 
Bethe Ansatz, the Truncated Conformal Space approach and 
perturbation theory (for an overview see e.g.\ 
\cite{Graham:2000si}). These techniques have helped to 
accumulate a rather extensive knowledge about boundary RG-flows 
in specific models. On the other hand, with the exception of the 
`g-conjecture' \cite{Affleck:1991tk}, we lack model independent 
principles that could guide us to predict possible boundary flows. 
Only in the case of WZW models, Affleck and Ludwig found an easy 
to formulate rule \cite{Affleck:1991by}. Here we shall present a 
generalization of their `absorption of the boundary spin'-principle 
which applies to all fixed-point free coset conformal field 
theories. 
\smallskip

Our presentation starts with a precise formulation of the new 
principle. It is then argued that our proposal is consistent 
with the perturbative results obtained in \cite{Fredenhagen:2001kw}. 
Finally, we apply the proposed rule to coset realizations of 
unitary minimal models and compare our `predictions' with  
known results. 
\medskip 

Let $G/H$ be a coset conformal theory with primaries labeled 
by pairs $(l,l')$ of integrable highest-weight representations 
$l$ and $l'$ of the affine Lie algebras $\hat{\mathfrak{g}}_{k}$ and 
$\hat{\mathfrak{h}}_{k'}$, respectively. Note that in many examples, 
branching selection rules restrict the admissible pairs and 
that different pairs may describe the same coset sector. 
For our purposes, however, there is no need to be more specific 
about such issues. Boundary conditions $(L,L')$ of so-called {\em 
Cardy type} \cite{Cardy:1989ir} can be labeled by elements 
from the same set, i.e.\ by primaries of $G/H$. It is natural 
to extend this correspondence between elementary Cardy type 
boundary conditions and primaries (or their conformal 
families) such that mixtures (or `superpositions') of 
boundary theories are associated with sums of 
conformal families.  

Let us now choose representations $\sigma ,L$ of $\hat{\mathfrak{g}}$ and 
$L'$ of $\hat{\mathfrak{h}}$. Then our rule predicts the following flow 
between boundary conditions, 
\begin{equation}\label{flow}
(L,\sigma |_{\mathfrak{h}}\; \hat{\times}\;  L')\ \longrightarrow (L \; 
\hat{\times} \; \sigma,L')\ .
\end{equation}
Here, $\hat{\times} $ denotes the fusion product for representations of 
the affine Lie algebras $\hat{\mathfrak{g}}$ and $\hat{\mathfrak{h}}$, respectively. 
The definition of the restriction in $\sigma |_{\mathfrak{h}}$ is not entirely 
obvious. It is based on restricting the corresponding representation 
$\sigma_0$ of the finite dimensional Lie algebra $\mathfrak{g}$ to its 
subalgebra $\mathfrak{h}$. The existence of an embedding between the two 
affine algebras guarantees that all the subrepresentations in 
$\sigma_0 |_{\mathfrak{h}}$ give rise to integrable highest-weight 
representations of $\hat {\mathfrak{h}}_{k'}$. Their sum is then denoted 
by $\sigma |_{\mathfrak{h}}$. Note furthermore that in most cases, the 
boundary labels on both sides of the flow (\ref{flow}) involve 
reducible representations. To identify the configurations as 
mixtures of elementary boundary conditions, we have to decompose 
the representations into irreducibles.

It is certainly of interest to specify which boundary field is 
responsible for the flow (\ref{flow}). Even though this issue can 
be analyzed in more detail, we shall content ourselves with some 
simple statements. They involve the integrable highest-weight 
representations $\theta$ and $\theta'$ which are built from the 
adjoint representations of the Lie algebras $\mathfrak{g}$ and $\mathfrak{h}$, 
respectively. Our main rule asserts that the flows (\ref{flow}) 
are generated by fields from the coset sectors 
\begin{equation} \label{list}  
\mathcal{H} _{(0,l')}\ \ , \ \ \mbox{ where } \ \ 
   l' \subset \theta |_{\mathfrak{h}} \ \ \  .   
\end{equation} 
Moreover, if the initial boundary theory before perturbation 
contains the sector $\mathcal{H}_{(0,\theta')}$ at most once, then 
$l' = \theta'$ can be omitted from the list (\ref{list}).   

Let us observe in passing that we recover the principle found by 
Affleck and Ludwig when we specialize to the example of WZW-models, 
i.e.\ to coset models with trivial denominator. Indeed, in this 
case the flow (\ref{flow}) reduces to
\[
\text{dim} (\sigma)\ (L)\ \longrightarrow \ (L \; \hat{\times}\;  \sigma) 
\ = \ {\bigoplus}_{J}\  {N_{\sigma L}}^{J}
\ (J)
\]
where dim$(\sigma)$ is the dimension of the representation 
$\sigma_0$ of $\mathfrak{g}$ and $N$ denote the fusion rules of the 
affine Lie algebra $\hat{\mathfrak{g}}$. The perturbing field is given 
by the product $S^{\sigma } J(x)$ which describes the coupling 
of the current $J(x)$ to some boundary spin $S^\sigma$. Comparison 
with \cite{Affleck:1991by} shows that we have reproduced the 'absorption 
of the boundary spin'-principle. 
\medskip 

Before we move on to examples and applications of the stated rule, 
we want to discuss its relation with the perturbative results 
obtained in \cite{Fredenhagen:2001kw}. There, we considered 
coset models in a limiting regime in which some of the involved 
levels become large. Using a perturbative approach we identified 
RG fixed points $Q$ in the vicinity of a chosen boundary condition 
$P$. To formulate the main result of \cite{Fredenhagen:2001kw}, let
us assign representations $P_0$ and $Q_0$ of the finite dimensional 
Lie algebra $\mathfrak{g} \oplus \mathfrak{h}$ to the boundary conditions $P$ and
$Q$ (see above). With this notation, we are able to state that 
$Q$ appears in the vicinity of $P$ if the associated representations
$P_0$ and $Q_0$ are equivalent on the diagonally embedded 
$\mathfrak{h}_{\rm diag} \subset \mathfrak{g} \oplus \mathfrak{h} $,
\begin{equation} \label{cond} 
P_0|_{\mathfrak{h}_{\rm diag}}\ \sim \ Q_0|_{\mathfrak{h}_{\rm diag}}\ \ .
\end{equation} 
Here, we have to assume that $P$ and $Q$ coincide in the directions 
in which the level is not sent to infinity (see \cite{Fredenhagen:2001kw}
for details). The boundary theories $Q$ satisfying the condition 
(\ref{cond}) provide candidates for the infrared fixed points of 
an RG flow which initiates from $P$. 

For comparison, let us now evaluate our rule (\ref{flow}) in the limiting 
regime, assuming that the representation $\sigma $ is trivial in 
the directions belonging to small levels. Under this condition, the 
fusion products in rel.\ (\ref{flow}) can be replaced by the usual 
tensor products of Lie algebra representations. Taking the 
configurations of both sides of (\ref{flow}) and restricting 
the associated representations to the diagonally embedded $\mathfrak{h}$, we 
obtain
\[
L|_{\mathfrak{h}}\otimes \sigma|_{\mathfrak{h}}\otimes L' \ \longrightarrow\ (L\otimes
\sigma )|_{\mathfrak{h}}\otimes L'\ .
\]
The two sides are equivalent because the decomposition of
representations commutes with taking tensor products.
\medskip 

As an application of our rule, let us consider the unitary minimal
models. They can be realized as diagonal coset models of the form 
$\text{su} (2)_{k}\oplus \text{su} (2)_{1}/\text{su} (2)_{k+1}$. 
Correspondingly, their sectors are labeled by three integers 
$(l,s,l')$ in the range $l=0\dots k$, $s=0,1$, $l'=0\dots k+1$. 
Branching selection rules restrict $l+s+l'$ to be even, and there 
is an identification  $(l,s,l')\sim (k-l,1-s,k+1-l')$ between 
admissible labels. Our rule (\ref{flow}) predicts flows for a 
large number of starting configurations. Many of them are 
superpositions of boundary conditions, but here we will concentrate 
on perturbations of a single boundary condition $(J,S,J')$. Let us 
assume that $1\leq J' \leq k$. Then we choose the representation
$\sigma$ of the numerator theory as $\sigma = (J',0)$ and fix $L'$ 
to be $L'=(0)$. With these choices our rule becomes 
\begin{equation}\label{pertflow}
(J,S,J')\ \longrightarrow\ \bigoplus_{L}\ 
{N_{J\, J'}}^{L}\ (L,S,0)\ \ 
\end{equation}
where $N$ denote the fusion rules of su(2)$_k$.
On the other hand, if we select $\sigma $ to be $(k+1-J',0)$ and 
$L' = (k+1)$, we find
\begin{equation}\label{nonpertflow}
(J,S,J')\ \longrightarrow\ \bigoplus_{L}\ 
{N_{J\,J'-1}}^{L}\ (L,1-S,0)\ \ .
\end{equation}
The first of these flows can be seen in perturbation theory for large
level $k$ \cite{Recknagel:2000ri,Graham:2001pp}, whereas the second 
does not become 'small' in this limit. Nevertheless, both flows are 
known to exist \cite{Chim:1996kf,Lesage:1998qf,Ahn:1998xm}. They 
are generated by the $(0,0,2)$ field (in standard Kac labels $(1,3)$) 
and differ by the sign of the perturbation. This is in agreement with 
our general statements on the boundary fields generating the flow 
(\ref{flow}). 
\smallskip

In the simplest minimal model, the critical Ising model, there are
three possible elementary boundary conditions: the free boundary 
condition $(0,1,1)$, and boundary conditions $(0,0,0),(1,1,0)$ in 
which the boundary spin is forced to be either up or down. Starting 
from the free condition, the system can be driven into a theory with 
fixed spin \cite{Ghoshal:1994tm}. These are precisely the two flows 
(\ref{pertflow}), (\ref{nonpertflow}).

The second model in the unitary minimal series is the tricritical Ising
model with central charge $c=7/10$. Once more, the flows triggered by 
the $\phi_{13}$ field \cite{Chim:1996kf} are correctly reproduced by 
(\ref{pertflow}) and (\ref{nonpertflow}). There are, however, more 
flows known which correspond to a perturbation with other fields 
\cite{Affleck:2000jv}. As our rule depends on the specific coset 
construction, it is possible to find additional flows by choosing  
different coset realizations of the same theory. For the tricritical 
Ising model, such an alternative realization does exist. It is 
given by $(\text{E}_{7})_{1}\oplus(\text{E}_{7})_{1}/(\text{E}_{7})_{2} $. 
When we apply our rule to this coset construction, it reproduces the two
known flows caused by the $\phi_{33}$ field. In Kac labels they read 
\[
(2,2)\ \longrightarrow\ (3,1)\ ,\ \ (2,2)\ \longrightarrow\ (1,1)\ .
\]
These two flows also appear in higher minimal models \cite{Graham:2001tg} 
where we do not know a coset realization for the
$\phi_{33}$-perturbations. 
This may be related to the observation that the tricritical Ising model 
seems to be the only theory in which the considered perturbations are 
integrable \cite{Graham:2001tg}. Nevertheless, recovering flows from
the exceptional E$_{7}$ coset construction can be considered as an
important check of the conjectured rule.
\medskip 

These examples may help to illustrate the wide applicability 
of our new rule. It is even possible to generalize the rule in a
straightforward way beyond the {\em Cardy case} when dealing with twisted
boundary conditions or with modular invariants not given by charge
conjugation. While further checks certainly remain to be 
done, we hope that our proposal provides an elegant way to 
summarize results obtained from RG computations and that it 
will emerge as simple guide to predicting new flows.  
\smallskip  

\noindent
\begin{acknowledgments} 
We would like to thank T.\ Quella for stimulating discussions. We are
also grateful to 
K.\ Graham, G.\ Watts and A.\ Zamolodchikov for their comments and a 
careful reading of the manuscript. 
\end{acknowledgments}

\end{document}